\documentclass[%
 reprint,
 amsmath,amssymb,
 aps,
]{revtex4-2}
\usepackage{dcolumn}
\usepackage{blindtext}
\usepackage[utf8]{inputenc}
\usepackage{amsmath}
\usepackage{amssymb}
\usepackage{amsthm}
\usepackage{setspace}
\usepackage{graphicx}
\usepackage{bm}
\usepackage{braket}
\usepackage{mathrsfs}
\usepackage{physics}
\usepackage{float}
\usepackage[colorlinks = true,linkcolor = red,urlcolor  = blue,citecolor = blue,anchorcolor = blue]{hyperref}
\usepackage[english]{babel}
\usepackage{bm}
\usepackage [autostyle, english = american]{csquotes}
\MakeOuterQuote{"}
\usepackage{pifont}
\begin{document}
\preprint{APS/123-QED}

\title{Controllable Single Photon Scattering via Coupling of Driven $\Lambda$ System with Topological Waveguide}
\author{Gunjan Yadav}
\author{Madan Mohan Mahana}
\author{Tarak Nath Dey}
\thanks{tarak.dey@iitg.ac.in}%
\affiliation{Department of Physics, Indian Institute of Technology Guwahati, North Guwahati 781039, Assam, India}

\begin{abstract}
We investigate the coherent single photon scattering process in a topological waveguide coupled with a driven $\Lambda$ system. We derive an analytical expression for transmittance by using the scattering formalism for three different sublattice sites (A, B, and AB), which couples to the $\Lambda$ system. We have demonstrated that the system's response is topology-independent for A and B sublattice-site coupling and becomes topology-dependent for AB sublattice-site coupling. In a weak control field regime, the system behaves as a perfect mirror in all of these configurations. Upon the control field strength enhancement, the transmission spectrum evolves from Electromagnetically Induced Transparency (EIT) to Autler-Townes splitting (ATS) in A and B sublattice-site coupling. The manipulation of transmission from opaque to transparent holds the key mechanism of a single photon switch. Further, the topology-dependent AB sublattice configuration allows the sharper Fano line shape that is absent in topology-independent  A and B sublattice configurations. This characteristic of the Fano line can be used as a tunable single-photon switch and for sensing external perturbations. Furthermore, our study paves the way for the robustness and tunability of systems with applications in quantum technologies such as quantum switches, sensors, and communication devices. 
\end{abstract}
\maketitle
\section{Introduction}
\label{Sec I}
Strong light-matter interactions and single photon scattering are at the heart of many emerging technologies in quantum information science, including quantum computing, secure communication, quantum simulation, and advanced sensing \cite{PhysRevLett.111.090502,Kannan2023,PhysRevLett.131.073602, Du2020ControllableOR}. Achieving efficient coupling between an emitter and a structured environment is crucial for realizing these applications. Over the past few decades, the coupling of emitters with various platforms, such as cavity structures, linear waveguides, and photonic crystals, has been studied quite well \cite{Mabuchi2002,PhysRevLett.113.093603}. Among these platforms, waveguide quantum electrodynamics (QED) is promising for enhancing light-atom interactions by confining light to a single dimension. Unlike cavity QED, which is limited to discrete modes, waveguide QED supports a continuum of modes, offering unique advantages \cite{roy2017colloquium}. Significant progress has been made in waveguide QED systems, including photonic crystal waveguides \cite{PhysRevLett.115.063601}, trapped atom coupled to an optical fiber \cite{PhysRevLett.117.133603}, and superconducting qubits coupled with transmission lines \cite{Astafiev2010, PhysRevLett.108.263601}. However, a key challenge in building practical quantum systems is maintaining robustness against imperfections and disorder. One promising approach is to combine topological concepts with quantum systems. This integration allows for more interesting features, which are hard to achieve in conventional systems.

Topology, originally a mathematical field focused on properties that remain invariant under continuous deformations,  such as stretching, twisting, or bending, without tearing or cutting. In recent decades, this abstract mathematical framework has found novel applications in physics, providing a new lens to study robust systems against local imperfections. The concept of topology first gained significant attention in condensed matter physics with the study of topological insulators \cite{PhysRevLett.45.494}. Following the realization of the photonic hall effect in photonic crystals, the concept of topology has been explored in photonic systems, resulting in advancements such as edge state lasing and unidirectional propagation of light 
\cite{PhysRevLett.101.100501,  Ota2018, PhysRevLett.100.013905, wang2009observation}. These phenomena arise due to topologically protected modes that are immune to imperfections, providing robust solutions for photonic technologies. The exploration of topology has since extended into quantum systems, where one of the most fundamental and widely studied models is the one-dimensional Su-Schrieffer-Heeger ( 1D SSH) model.  This model is particularly valuable because it captures essential topological properties in a relatively simple system. It has been employed in various photonic and quantum systems to investigate a variety of optical phenomena, including quantum sensing \cite{Zhang2023}, edge state lasing \cite{st2017lasing}, harmonic generation \cite{yuan2022giant}, topological protection of biphoton states \cite{blanco2018topological}, and the generation of topologically protected entangled states \cite{Wang2019TopologicallyPE}.

The interaction between a two-level quantum emitter and a topological waveguide has recently attracted significant attention, with experimental realizations demonstrated in superconducting circuits \cite{bello2019unconventional,PhysRevX.11.011015}. The emitter couples to the band gap region of the topological waveguide's energy spectrum, giving rise to a directional bound state. In systems with multiple emitters, they interact via the overlap of these bound-state wavefunctions, leading to collective effects \cite{PhysRevA.104.053522,yadav2022bound, PhysRevX.10.031011}. In contrast, the emitter couples to the band pass region of the topological waveguide spectrum giving rise to single photon scattering. The coupling of multiple emitters at specific positions results in coherent emission, forming super-radiant and sub-radiant phenomena \cite{Tiranov2023,PhysRevA.89.013805,PhysRevX.11.011015,PhysRevA.97.043831}.\

In this work, we explore the single photon scattering problem, based on the theoretical scattering formalism approach. The motivation of this work is to find controllable single photon scattering and enable tunable transmission. We proposed an array of coupled resonators (A and B), which provide a photon transport channel in 1D continuum. The $\Lambda$ system can be placed in three distinct possible configurations, based on configuration it can show topology-independent and topology-dependent scattering. In the absence or presence of a control field, it can either reflect or transmit the single photon because of the absence or presence of the quantum interface effect in the system. The topological waveguide shows the topological properties in AB configuration, single photon scattering can take advantage of it for robust propagation of photons through 1D channel, and may find novel applications in quantum technologies. The coupling of $\Lambda$ system with a 1D linear waveguide has been broadly studied for applications in photon scattering \cite{PhysRevLett.106.053601}. In comparison to linear waveguide design in which the photon travels continuously, our proposed cavity array topological waveguide lets the photon travel discretely through a 1D channel by locally annihilating or creating a photon within the cavity. It also provides a rich energy band spectrum with a nonlinear dispersion relation, as opposed to the conventional linear dispersion relations. \

We present a model consisting of a topological waveguide coupled with a $\Lambda$ system and derive the transmittance expression. It shows the topology-independent and topology-dependent scattering in A, B, and AB configurations respectively. In the regime of a weak control field, the system effectively acts as a perfect mirror for photon scattering. As the control field strength increases, the transmission spectrum shifts from EIT to the ATS regime. Furthermore, we find a sharp and robust Topological
 Fano line shape in the AB configuration.

This manuscript is organized as follows: In Sec. \ref{sec:II}, we provide a detailed description of the topological waveguide and the 
$\Lambda$ system. In Sec. \ref{sec:III}, we discuss the single-photon excitation eigenstate, which is essential for studying single-photon scattering. We then explore the different sublattice-site couplings and their transfer matrices. The experimental realization of the setup is presented in Sec. \ref{sec:IV}. Finally, we conclude with a summary and outlook in Sec. \ref{sec:V}.

\section{Model}
\label{sec:II}
The 1D SSH model is fundamental in condensed matter physics. It provides a simple and powerful framework for exploring topological phases in physical systems. It has been adapted to explore various optical phenomena in photonics, with its applications extending to photonic counterparts, such as the 1D topological waveguide \cite{bello2019unconventional}. This topologically protected waveguide presents unique advantages over conventional types as discussed in Sec. \ref{Sec I}.  Coupling a $\Lambda$ system with topological waveguides allows for controllable single-photon scattering, offering promising opportunities for advancements in quantum information processing, efficient photon switching, and secure quantum communication systems. 

In this section, we present our model. In Sec. \ref{sec:IIA}, we explain the Hamiltonian, energy band structure, and the topological properties of the waveguide. Sec. \ref{sec:IIB} focuses on the details of the $\Lambda$ system and its Hamiltonian.
\subsection{Topological waveguide}
\label{sec:IIA}
\renewcommand{\figurename}{FIG.}
\begin{figure}
  \centering
\includegraphics[width= 0.9\linewidth]{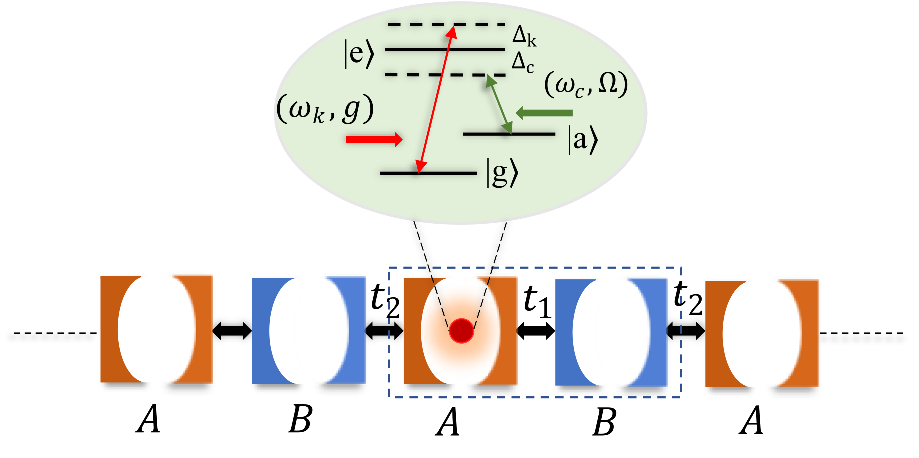}
 \caption{A schematic of the model system is shown. The topological waveguide is made up of an array of coupled resonators, labeled A (in orange) and B (in blue).  The sublattice-sites A and B together represent a unit cell of an array, which is highlighted with a dotted black box. The black double arrow represents intracell and intercell hopping parameters $t_1 = J(1+\delta)$ and $t_2 = J(1-
 \delta)$ respectively, where $\delta$ is the dimerization constant and $J$ is the characteristics energy parameter. The transition $\ket{g} \xrightarrow{} \ket{e}$ in the $\Lambda$ system is coupled to the topological waveguide mode with a coupling constant $g$, represented by the solid red line.  The transition from $\ket{a} \xrightarrow{} \ket{e}$ is driven by a classical control field with frequency $\omega_c$ and Rabi frequency $\Omega$, shown with a green arrow.}
 \label{Figure 1}
\end{figure}
 The topological waveguide consists of alternating coupled resonators, A and B, where each pair (A and B) forms a unit cell, as shown in Fig. \ref{Figure 1}. 
These resonator interactions, within inter and intracells are arise from evanescent-field coupling. The Hamiltonian of the 1D topological waveguide $(H_{TW})$ for $N$ unit cells, within the tight-binding approximation, can be expressed as
\begin{equation}
    H_{TW} = \sum_{j=1}^N \omega_0(a_j^{\dagger}a_j + b_j^{\dagger} b_j) -(t_1 a_j^{\dagger}b_j + t_2 a_{j+1}^{\dagger}b_j + \text{h.c}). 
    \label{Equation (1)}
\end{equation}
Here, $\omega_0$ represents the on-site energy, which is taken as the reference energy of the system. The parameters $t_1$ and $t_2$ denote the intracell and intercell hopping strengths, respectively, and are related to the dimerization constant $\delta$ as $t_1 = J(1 + \delta)$ and $t_2 = J(1 - \delta)$. The parameter $J$ is a characteristic energy parameter or average hopping strength, it sets the scale for the hopping interactions in the system. The dimerization constant $\delta$, represents the relative strength of intracell and intercell hopping in a system with a periodic lattice structure. In such systems, the hopping between neighboring sites be alternates, and leads to the formation of dimers such as pairs of sites with stronger or weaker bonds, depending on $\delta$. For $\delta = 0$, the system has uniform hopping with no dimerization ($t_1 = t_2 = J)$. For $\delta > 0$, the intracell hopping is stronger ($t_1 > t_2)$, and for $\delta < 0$, the intercell hopping is stronger $(t_2 > t_1)$. This variation in hopping leads to different topological properties in the system \cite{Asboth2016}.
The operators $a_j^{\dagger}$ ($a_j$) and $b_j^{\dagger}$ ($b_j$) represent the creation (annihilation) operators of a photon at the A and B sublattice-sites in the $j^{th}$ unit cell. For simplicity, we assume a low-loss waveguide such that photonic loss from each cavity can be neglected. The Hamiltonian provided in Eq. (\ref{Equation (1)}) is formulated in real space for a finite $N$ number of unit cells. By applying periodic boundary conditions, this Hamiltonian can be transformed into momentum space using the discrete Fourier transform of the operators as
\begin{equation}
    a_k = \frac{1}{\sqrt{N}} \sum_j {e^{-i k j}a_j}, \hspace{0.2cm}  b_k = \frac{1}{\sqrt{N}} \sum_j {e^{-i kj}b_j}.
\end{equation}
\noindent After performing the transformation, the Hamiltonian in momentum space can be expressed as follows: $H_{TW}  = \sum_k V_k^{\dagger} H_k V_k$, with (setting $\hbar = 1$)
\begin{equation}
    V_k = \begin{pmatrix}
        a_j \\
        b_j
    \end{pmatrix},
    \quad
     H_k =
     \begin{pmatrix}
        \omega_0 & h(k) \\
        h^*(k) & \omega_0
    \end{pmatrix}.
    \label{Eq.3}
\end{equation}
\noindent Here, $h(k) = -t_1 - t_2 e^{-i k} = |h(k)| e^{i \phi_k}$ defines the coupling between A and B modes, with $\phi_k$ representing the phase of the system, given as $\phi_k = $ arg$(h(k)).$
For further calculations, we set $\omega_0$ to be zero to preserve the chiral symmetry of the structure, $\sigma_z H_k \sigma_z = - H_k$ \cite{bello2019unconventional}. Diagonalizing the matrix $H_k$ provides the system's eigenvalues and eigenvectors, which describe the energy-momentum dispersion relation and eigenbands of the topological waveguide. The eigenvalue expression is calculated as
\begin{equation}
    \omega_k = |h(k)| = \pm\sqrt{t_1^2 + t_2^2 + 2t_1t_2\cos{(k)}}.
\end{equation}
The spectrum consists of two bands: the upper band ($\omega_k$) and the lower band ($-\omega_k$), spanning a range $[2|\delta|J, 2J]$, and $[-2J, -2|\delta|J]$ respectively, as illustrated in Fig. \ref{fig:combined}(a). Photons can propagate within these bands with a velocity given by $v_g = \partial \omega_k/ \partial k$. The gap between the upper and lower bands, which spans $4|\delta|J$, is referred to as the bandgap. In this bandgap, photons are unable to propagate through the waveguide. The eigenvectors associated with these upper and lower bands are
\begin{equation}
    u_k/l_k = (\pm a_k + e^{i \phi} b_k)/\sqrt{2}.
\end{equation}
The Hamiltonian of the topological waveguide in momentum space can also be expressed in terms of these eigenvalues and eigenvectors as: $H_B = \sum_k \omega_k (u_k^{\dagger} u_k - l_k^{\dagger} l_k)$.
\renewcommand{\figurename}{FIG.}
\begin{figure}[H]
  \centering
\includegraphics[width= 1\linewidth]{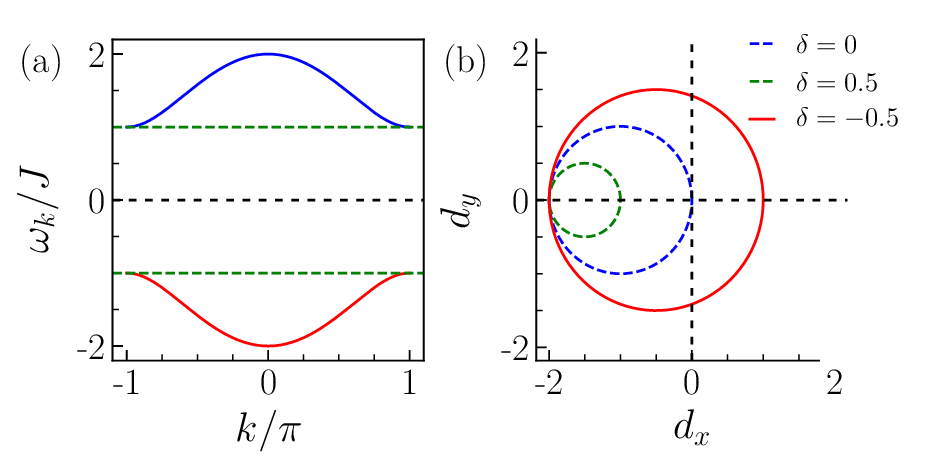}
\caption{(a) Energy-momentum dispersion of the topological waveguide with a dimerization parameter $|\delta| = 0.5$.  The blue curve represents the upper band, while the red curve shows the lower band, corresponding to energies $\omega_k$ and $-\omega_k$, respectively. The area between the two green dotted lines marks the bandgap. (b) The trajectory of the vector ${d}(k) = (d_x(k), d_y(k))$ is shown for different values of the dimerization parameter $\delta$. The solid red, dotted blue, and dotted green curves correspond to $\delta = -0.5$, $\delta = 0$, and $\delta = 0.5$, respectively.}
\label{fig:combined}
\end{figure}
The topological waveguide can exist in two distinct phases: topological and trivial. These phases are determined by the sign of the dimerization constant, $\delta$. Although the band spectrum looks identical for both $\delta > 0$ and $\delta < 0$ (since the expression for $\omega_k$ does not depend on the sign of $\delta$), these represent fundamentally different topological phases.  The transition between the two phases is marked by the closing and reopening of the band gap, which signifies a topological phase transition. Changing the sign of $\delta$ evolves the system from one phase to the other \cite{Asboth2016}.
 To fully characterize and differentiate these two cases, we adopt an alternative approach. Since the Hamiltonian $H_k$ is a $2 \times 2$ matrix, it can be expanded in terms of the Pauli matrices $\sigma_{\alpha}(\alpha \in x,y,z)$ as follows
 \begin{equation}
     H_k = \sum_{\alpha = x,y,z}d_{\alpha}(k)\sigma_{\alpha},
 \end{equation}
where $\sigma_x, \sigma_y$ and $\sigma_z$ are the Pauli matrices, and the vector $d_{\alpha}(k)$ represents the $x,y$ and $z$ component of pseudospin vector $d(k)$, which depend on momentum $k$ and describe the system's dynamics in reciprocal space. By comparing this expression with the form of $H_k$ from Eq. (\ref{Eq.3}), we obtain $h(k)= -t_1 - t_2 \cos(k) + it_2 \sin (k) =  d_x(k) - i d_y (k)$. The $k$-dependent components of the three-dimensional vector $d(k)$, reads
\begin{equation}
        d_x  = -t_1 - t_2 \cos{(k)},\hspace{0.4cm} 
        d_y = -t_2 \sin{(k)}, \hspace{0.4cm}
         d_z = 0.
\end{equation}
Here, $d_x$ is the Re$[h(k)]$ and $d_y$ is the Im$[h(k)]$. This formulation provides a clearer understanding of how the system behaves as the dimerization constant $\delta$ changes.
The trajectory of $d_x$-$d_y$ has been plotted for three different values of $\delta$, as shown in Fig. \ref{fig:combined}(b). This trajectory helps us to define a topological invariant known as the winding number $\nu$, which counts how many times the vector ${d}(k) = (d_x(k), d_y(k))$ winds around the origin as $k$ traverses the Brillouin Zone (BZ). For $\delta = 0$, as shown by the dotted blue line, indicating a closed band gap where $\nu$ is undefined. For $\delta > 0$, the loop does not enclose the origin, shown in dotted green, corresponding to $\nu = 0$. For $\delta < 0$, the loop encloses the origin, represented by the solid red line, corresponding to $\nu = 1$. Following Berry's seminal work, the Berry phase describes a geometric phase acquired in two-dimensional systems \cite{PhysRevLett.62.2747}. In 1D systems, this corresponds to the Zak phase,  which is related to the winding number \( \nu \) by the formula: $
\theta_{\text{Zak}} = \nu \pi$.
 Thus, for $\delta > 0 (t_1 > t_2)$, the Zak phase is $0$, corresponding to a trivial phase. For $\delta < 0 ( t_1 < t_2)$, the Zak phase is $\pi$, indicating a topologically non-trivial phase \cite{PhysRevX.11.011015}. In a physical system (\textit{i.e.,} in the bulk), the role of $\delta$ can be reversed by shifting the unit cell.
\subsection{Driven $\Lambda$ System}
\label{sec:IIB}
After understanding the topological properties of the waveguide and its phases, we now move on to the driven $\Lambda$ system. The topological waveguide described earlier is coupled with a driven $\Lambda$ system at a specific sublattice-site. The coupling of the $\Lambda$ system with A sublattice site is shown in Fig. \ref{Figure 1}. This configuration can be modified to the B sublattice-site coupling by relocating the $\Lambda$ system from A to the B sublattice-site. Additionally, an AB configuration can be achieved by positioning the $\Lambda$ system near the waveguide, between the A and B resonators. The $\Lambda$ system consists of three - levels: ground state $\ket{g}$, metastable state $\ket{a}$ and excited state $\ket{e}$. The dipole allowed transitions are $\ket{g} \xrightarrow{} \ket{e}$ and $\ket{a} \xrightarrow{} \ket{e}$. We consider that the transition $\ket{g} \xrightarrow{} \ket{e}$ is coupled with a photonic mode of the topological waveguide, while the transition from $\ket{a} \xrightarrow{} \ket{e}$ is driven by a classical control field. The bare Hamiltonian ($H_A$) of the $\Lambda$ system in the presence of the control field can be expressed as
\begin{equation}
    H_A = \sum_{i = e,a}\omega_i\ket{i}\bra{i} + \frac{\Omega}{2}(\ket{e}\bra{a} + \ket{a}\bra{e}).
\end{equation}
Here, the energy of the ground state $\ket{g}$ is set as the reference energy of the $\Lambda$ system. The energies $\omega_e = \omega_{eg}$ and $\omega_a = \omega_e-\Delta_c$ correspond to the excited state $\ket{e}$ and the metastable state $\ket{a}$, respectively \cite{PhysRevLett.107.223601}. The term $\omega_{eg}$ is the resonant frequency for the transition $\ket{g} \xrightarrow{} \ket{e}$, $\Delta_c$ is the detuning of the control field for the transition $\ket{a} \xrightarrow{} \ket{e}$, and $\Omega$ is the Rabi frequency of the control field. Here, we assume that there is no atomic loss (spontaneous emission) to the channel other than the waveguide. If there were, we would need to include a $i\Gamma/2$ term to represent the decay of the levels into other channels. The interaction of the $\Lambda$ system with the waveguide depends on the sublattice-site it is coupled to, and its interaction Hamiltonian will be modified accordingly, which we will discuss in the next section.\

\section{Single-Photon Scattering: \newline $\Lambda$ System Coupled to Topological Waveguide Bands}
\label{sec:III}
The two-band dispersion of a topological waveguide allows us to tune the transition energy $\omega_{eg}$ of $\Lambda$ system in two regions, \textit{\textit{i.e.}}, the bandpass and band-gap region. The transition energy lies within the band gap leading to the formation of atom-photon bound states \cite{PhysRevA.104.053522}.  Conversely, when it resonant with the bandpass region (upper and lower band) leads to scattering phenomena \cite{PhysRevA.109.063708}. We are interested in studying single photon scattering, hence our region of interest is the bandpass region.\

In this section, we explore single photon scattering in various configurations. In Sec. \ref{sec:IIIA} we discuss the single photon excitation eigenstate. In Sec. \ref{sec:IIIB}, \ref{sec:IIIC}, and \ref{sec:IIID} we discuss the single photon scattering for $\Lambda$ system coupled to a topological waveguide in A, B, and AB sublattice-site configuration, respectively.
\subsection{Single photon excitation}
\label{sec:IIIA}
To study photon transport in our system, we consider the single-excitation subspace, which is conserved by the total Hamiltonian, \textit{i.e.}, $[H, \hat{n}_e + \hat{n}_{atom}] = 0$. This implies that, within this subspace, only one excitation can exist at a time, meaning either the waveguide or the atom is excited, but not both simultaneously. We consider an incident photon as a plane wave, following a similar approach to that used for a two-level system \cite{bello2019unconventional}. The photon incident from the left end of the waveguide is resonant with the bandpass region of the waveguide's dispersion. As the photon interacts with the system, it gets scattered. An ansatz for the wavefunction in the single-excitation subspace can be written as
\begin{equation}
\begin{split}
     \ket{E_k} = & \left[ u_e \ket{e} \bra{g} + u_a \ket{a} \bra{g} \right. \\
     & \left. + \sum_j \left( u_A(j) a_j^{\dagger} + u_B(j) b_j^{\dagger} \right) \right] \ket{0,g}.
\end{split}
\end{equation}

Here, $u_A(j)$ and $u_B(j)$ represent the probability amplitudes
of finding a photonic excitation in A and
B sublattice-site of $j^{th}$ unit cell respectively. The term $u_e$ and $u_a$ correspond to the excitation amplitudes for the transitions $\ket{g} \xrightarrow{} \ket{e}$ and $\ket{g} \xrightarrow{} \ket{a}$, respectively. The scattering properties of a single photon interacting with a $\Lambda$ system initially in the ground state can be derived from the scattering eigenstates, which are solutions to the stationary Schrödinger equation (also known as the secular equation) as: $H\ket{E_k} = \pm \omega_k \ket{E_k}$. The scattering process results in either the transmission or reflection of the photon and depends on which sublattice-site the $\Lambda$ system is coupled to. In the following sections, we will develop the transfer matrix and analyze the transmittance for different sublattice-site couplings. \
\subsection{A Sublattice-site coupling}
\label{sec:IIIB}
In this configuration, we consider the $\Lambda$ system is coupled with only sublattice-site A at $x_1$ unit cell of the topological waveguide and probing the upper band ($\omega_k$) of the topological waveguide's dispersion. The transition $\ket{g} \xrightarrow{} \ket{e}$ of $\Lambda$ system is resonant with the upper passband energy of the topological waveguide, ensuring efficient interaction between the $\Lambda$ system and the topological waveguide modes. The interaction Hamiltonian $H_I^A$ of this configuration can be written as
\begin{equation}
    H_I^A = g(a_{x_1}^{\dagger}\ket{g}\bra{e} + \text{h.c}),
\end{equation}
where $g$ is the coupling constant of photonic waveguide mode with atomic transition, and $a_{x_1}^\dagger$ represents the creation operator of a photon at A sublattice-site in $x_1$ unit cell of the topological waveguide. The total Hamiltonian of the system is given by: $H_T^A = H_{TW} + H_A + H_I^A$.
By using the secular equation: $H_T^A\ket{E_k} = \omega_k\ket{E_k}$, the equation of motion is obtained, providing a framework to analyze the photon scattering in this configuration
\begin{subequations}
\renewcommand{\theequation}{\theparentequation\alph{equation}}
\begin{align}
\label{eq:11a}
(\omega_e-\Delta_c)u_a + ({\Omega}/{2})u_e = \omega_k u_a, \\
\label{eq:11b}
gu_A(x_1) + \omega_eu_e + ({\Omega}/{2})u_a = \omega_ku_e, \\
\label{eq: (11c)}
 -t_1u_B(x_1) - t_2u_B(x_1-1) + gu_e = \omega_ku_A(x_1).
\end{align}
\end{subequations}
By eliminating the probability amplitudes of the excited and metastable states and substituting them into Eq. (\ref{eq: (11c)}), we derive the scattering equation governing the coherent transport of a single photon as
\begin{equation}
    -t_1 u_B(x_1) - t_2 u_B(x_1-1) = (\omega_k - V) u_A(x_1), 
    \label{Eq. (12)}
\end{equation}
where $V$ represents the effective potential induced by the $\Lambda$ system at the coupling site and is given by
\begin{equation}
    V = \frac{4g^2(\omega_k-\omega_e+\Delta_c)}{4(\omega_k-\omega_e)(\omega_k-\omega_e+\Delta_c) - \Omega^2}.
    \label{Eq. (13)}
\end{equation}
Here, $\omega_k - \omega_e = \Delta_k$ denotes the detuning of the incident photon from resonant transition energy. The potential $V$ is observed to be dependent on the detuning of the incident photon $\Delta_k$, the detuning parameter of the control field $\Delta_c$, and the strength of Rabi frequency $\Omega$ associated with the control field. In the absence of $\Omega$, the effective potential reduces to $V = g^2/\Delta_k$. Any parameter variation alters the potential and consequently affects the single photon scattering. \
To solve the single photon scattering equation defined in Eq. (\ref{Eq. (12)}), we adopt the plane wave ansatz for the incident wave, as detailed in the Appendix. Applying the boundary condition at the coupling site, $u_A^-(x_1) = u_A^+(x_1)$, we derive the transfer matrix corresponding to the coupling at the A sublattice-site as
\begin{equation}
 U_A =
    \begin{pmatrix}
        1+\frac{V}{2 i t_2 \sin{(k+\phi)}} & \frac{V e^{-2i(kx_1+\phi)}}{2 i t_2 \sin{(k+\phi)}}\\
        -\frac{V e^{2i(kx_1+\phi)}}{2i t_2 \sin{(k+\phi)}} & 1-\frac{V}{2 i t_2 \sin{(k+\phi)}}
    \end{pmatrix}.
\end{equation}
Here, $U_A$ represents the transfer matrix for this configuration. From the scattering matrix, as discussed in the Appendix, the expression for transmittance is obtained as
\begin{equation}
    t_A = \frac{2 t_1 t_2 \sin{(k)}}{2 t_1 t_2 \sin{(k)} - i V \omega_k}.
    \label{Eq. (15)}
\end{equation}
 The calculated transmittance depends on incident energy resonant with the upper band ($\omega_k$), potential ($V$), and product of $t_1 = J(1+\delta)$ and $t_2 = J(1-\delta)$, which is independent of the sign of $\delta$.  
\subsection{B Sublattice-site coupling}
\label{sec:IIIC}
In this configuration, we mainly focus on the second sublattice-site within the $x_1$ unit cell of the waveguide \textit{i.e.}, the B sublattice-site. Here, the $\Lambda$ system is positioned near the B sublattice-site and the coupling takes place specifically at this sublattice-site. The interaction Hamiltonian for this configuration is defined as follows
\begin{equation} 
H_I^B = g(b_{x_1}^{\dagger}\ket{g}\bra{e} + \text{h.c.}). \end{equation} 
Here, $g$ is the coupling parameter, and $b_{x_1}^\dagger$ is the creation operator of a photon at B sublattice-site. The total Hamiltonian of the system is expressed as: $H_T^B = H_{TW} + H_A + H_I^B$. Again, by using the secular equation: $H_T^B \ket{E_k} =  \omega_k \ket{E_k}$, the equations of motion for this configuration are obtained as
\begin{subequations}
\renewcommand{\theequation}{\theparentequation\alph{equation}}
\begin{align}
\label{eq:11a}
(\omega_e-\Delta_c)u_a + (\Omega / 2) u_e = \omega_k u_a, \\
\label{eq:11b}
gu_B(x_1) + \omega_e u_e + ({\Omega}/{2})u_a  = \omega_k u_e, \\
\label{eq:11c}
-t_1u_A(x_1) - t_2u_A(x_1+1) + gu_e = \omega_k u_B(x_1).
\end{align}
\end{subequations}
Solving these equations of motion yields the scattering equation for this configuration
\begin{equation}
    -t_1u_A(x_1) - t_2u_A(x_1+1) = (\omega_k-V)u_B(x_1).
\end{equation}
Here, $V$ is the effective potential created by the $\Lambda$ system at B sublattice-site. The expression for $V$ is identical to that given in Eq. (\ref{Eq. (13)}). Using the boundary condition at the coupling position, $u_B^-(x_1) = u_B^+(x_1)$, and solving the scattering equation in the same manner as outlined in Sec. \ref{sec:IIIB}, we obtain the transfer matrix for the B sublattice-site coupling as
\begin{equation}
U_B = 
    \begin{pmatrix}
        1-\frac{V}{2i t_1 \sin{\phi}} & -\frac{V e^{-2i kx_1}}{2i t_1 \sin{\phi}}\\
        \frac{V e^{2i kx_1}}{2i t_1 \sin{\phi}} & 1+\frac{V}{2i t_1 \sin{\phi}}
    \end{pmatrix}.
\end{equation}
Here, $U_B$ represents the transfer matrix. The transmittance of this configuration is calculated as
\begin{equation}
    t_B = \frac{2 t_1 t_2 \sin{(k)}}{2 t_1 t_2 \sin{(k)} - i V \omega_k} .
    \label{Eq. (20)}
\end{equation}
\renewcommand{\figurename}{FIG.}
\begin{figure}
  \centering
        \includegraphics[width= 1\linewidth]{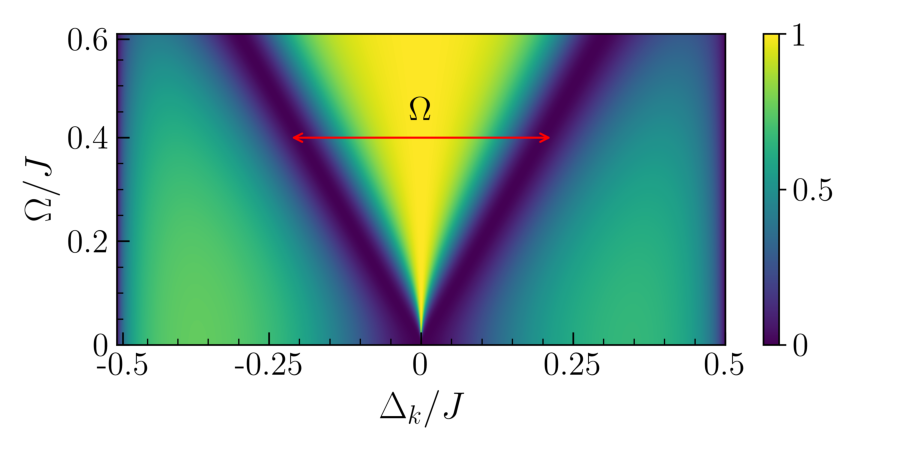}
 \caption{The 2D contour plot shows the transmission intensity for A sublattice-site coupling, $T_{A} = |t_{A}|^2$, as a function of incident photon detuning $\Delta_k/J = (\omega_k - \omega_e)/J$, and the strength of control field Rabi frequency $\Omega/ J$. The coupling constant is set to $g = 0.2J$, and atomic transition frequency $\omega_{eg} = \omega_e = 1.5J$ is fixed. Since the Eqs. (\ref{Eq. (15)}) and (\ref{Eq. (20)}) are identical, therefore this plot also hold good for B sublattice-site coupling }
 \label{fig. 3}
\end{figure}
The expressions for transmittance in both configurations, A and B, are identical, and the dimerization constant $\delta$ does not appear explicitly in Eqs. (\ref{Eq. (15)}) and (\ref{Eq. (20)}). It indicates that neither configuration is sensitive to the topological properties of the topological waveguide. Instead, the transmittance depends on the coupling strength $g$, the detuning $\Delta_c$, and the control field strength $\Omega$. This dependency allows for precise control of the transmittance by adjusting any of these parameters. 
Fig. \ref{fig. 3} illustrates the variation in transmission intensities $T_{A} = |t_{A}|^2$ and $T_{B} = |t_{B}|^2$, as a function of the incident photon detuning $\Delta_k/ J$ and the control field strength $\Omega/J$. Since the transmittance is identical for both configurations hence, $|t_{A}|^2 = |t_{B}|^2$. Notably, the yellow regimes in the contour plot represent the maximum transmission intensity, corresponding to a peak in the transmission spectrum. These peaks indicate the transparency window, characterized by a maximum transmission, and can be actively controlled by varying the strength of \(\Omega\). This tunability of the transparency window offers valuable insights into the system's behavior under different \(\Omega\) strengths and provides a flexible mechanism for modulating photon transmission in the setup.\

The observed variation in transmittance arises from the distinct behavior of the \(\Lambda\) system under different values of \(\Omega\), which can be explained as follows. In the absence of a control field, the \(\Lambda\) system behaves like a two-level system, acting as a perfect mirror when perfectly resonant with the incident photon, thereby reflecting the photon completely. However, in the presence of a nonzero control field, atomic coherences are induced between the different transition pathways of the \(\Lambda\) system, leading to quantum interference effects. Specifically, the atom in the ground state \(\ket{g}\) can be excited to the excited state \(\ket{e}\) via two possible pathways. The direct transition \(\ket{g} \rightarrow \ket{e}\), and the indirect transition through the intermediate state \(\ket{a}\), \textit{i.e.}, \(\ket{g} \rightarrow \ket{a} \rightarrow \ket{e}\). The destructive interference between these two quantum pathways results in zero absorption of the incident photon, allowing it to pass through without reflection, a phenomenon known as EIT. Importantly, the width of this transparency window is controlled by the magnitude of the Rabi frequency \(\Omega\). As \(\Omega\) increases, the transparency window broadens, eventually leading to the emergence of ATS in the large \(\Omega\) limit. This splitting further enhances the control over the transmission characteristics of the system, enabling the modulation of photon transmission across a wider range of detuning.\

The distinct behaviors in single photon scattering can also be understood by analyzing the poles of the transmittance. At the control field resonance (\(\Delta_c = 0\)), the poles for the incident photon detuning (\(\Delta_k\)) are given by
\begin{equation}
    \Delta_k = i \frac{g^2 \omega_k}{4 t_1 t_2 \sin{(k)}} \pm \sqrt{\frac{\Omega^2}{4} - \frac{g^4 \omega_k^2}{16 t_1^2 t_2^2 \sin^2{(k)}}}.
\end{equation}
Different control field strengths lead to distinct transmission behaviors: Weak control field strength (\(|\Omega| \ll |g^2 \omega_k / 2 t_1 t_2 \sin(k)|\)): One pole is purely imaginary and another is zero, resulting in a single dip at resonance with a Lorentzian lineshape, characteristic of strong reflection.
Intermediate control field strength (\(|\Omega| \approx |g^2 \omega_k / 2 t_1 t_2 \sin(k)|\)): The pole remains purely imaginary, leading to sharp EIT peaks.
Strong control field strength (\(|\Omega| \gg |g^2 \omega_k / 2 t_1 t_2 \sin(k)|\)): The poles develop real components, resulting in two distinct dips at \(\pm \Omega/2\), representing ATS with a characteristic doublet in the transmission spectrum.\
\renewcommand{\figurename}{FIG.}
\begin{figure}
    \centering
        \includegraphics[width= 1\linewidth]{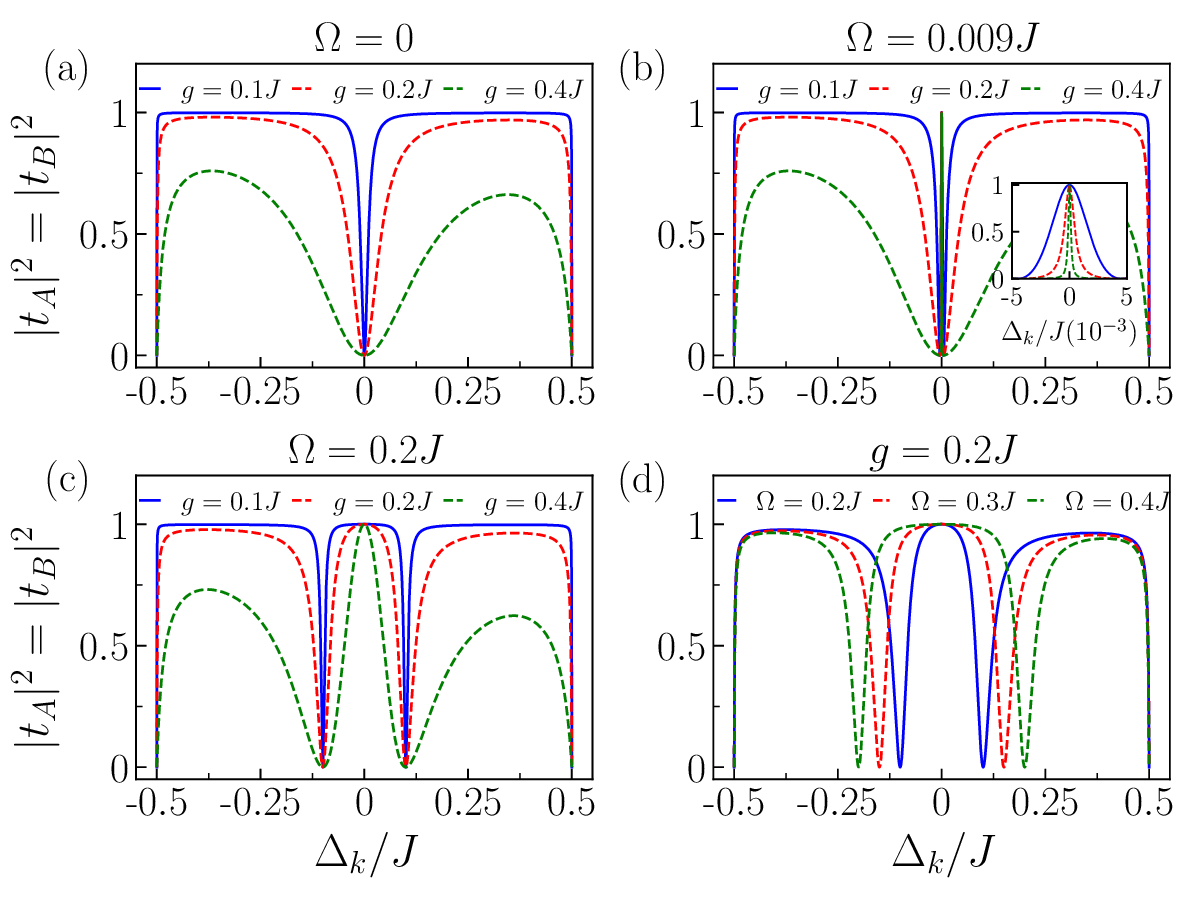}
 \caption{Transmission spectrum $T_{A} = |t_{A}|^2$, $T_{B} = |t_{B}|^2$ for the A and B sublattice-site coupling as a function of the incident photon detuning $ \Delta_k/J= (\omega_k - \omega_e)/J$. The atomic transition frequency is set to $\omega_e = 1.5J$. (a) In the absence of a control field \(\Omega = 0\). (b) In the presence of a control field with strength \(\Omega = 0.009J\).  The inset displays a magnified view of the EIT regime, providing a clearer visualization of its detailed features. (c) In the presence of a control field with strength \(\Omega = 0.2J\), the solid blue, dotted red, and dotted green color lines represent \(g = 0.1J\), \(0.2J\), and \(0.4J\), respectively. (d) Fixing \(g = 0.2J\) and varying the control field strength \(\Omega\) from $0.2J$ to $0.3J$ and $0.4J$, shown with solid blue, dotted red, and dotted green lines, respectively.}
 \label{fig. transmissionA}
\end{figure}
The transmission intensities, $|t_A|^2$ and $|t_B|^2$, are plotted in Fig. \ref{fig. transmissionA} as a function of the incident photon detuning $\Delta_k/J = (\omega_k - \omega_e)/J$, for various values of the coupling strength \(g\) and the control field Rabi frequency \(\Omega\). Fig. \ref{fig. transmissionA}(a) illustrates the variation of $|t_{A}|^2$ and $|t_{B}|^2$ in the absence of the control field (\(\Omega = 0\)). In this case, the transmission spectrum exhibit a Lorentzian lineshape, with a dip at the resonance frequency (\(\Delta_k = 0\)) as expected. \ 

Fig. \ref{fig. transmissionA}(b) presents the transmission spectrum for \(\Omega = 0.009J\), showing the emergence of sharp EIT peak.
In Fig. \ref{fig. transmissionA}(c), the transmission spectrum shows both the EIT window and ATS as \(\Omega\) is further increased. The transparency window, initially dominated by EIT, gradually transitions into the ATS regime as the control field becomes stronger. The appearance of ATS is characterized by the splitting of the transmission peaks, indicating a shift from quantum interference-based transparency to a regime dominated by the strong control field’s influence.
The colors in Figs. \ref{fig. transmissionA}(a), \ref{fig. transmissionA}(b), and \ref{fig. transmissionA}(c) represent different values of the coupling strength \(g\), where the solid blue, dotted red, and dotted green lines correspond to \(g = 0.1J\), \(g = 0.2J\), and \(g = 0.4J\), respectively.

Fig. \ref{fig. transmissionA}(d) focuses on the ATS behavior as a function of detuning. The transmission spectrum is plotted for different values of  $\Omega$. The solid blue, dotted red, and dotted green lines correspond to \(\Omega = 0.2J\), \(\Omega = 0.3J\), and \(\Omega = 0.4J\), respectively. As \(\Omega\) increases, the splitting between the transmission peaks becomes more pronounced, clearly illustrating the role of \(\Omega\) in controlling the ATS. This plot highlights the tunability of the transmission properties via the control field.\

The results presented in Figs. \ref{fig. transmissionA}(a)-\ref{fig. transmissionA}(d) demonstrate a tunable transmission behavior for single photon scattering through a system where both the atom-photon coupling strength $g$ and the control field strength $\Omega$ play critical roles. This tunability opens up potential applications in quantum photonic devices, particularly as a single photon switching device.
\subsection{AB Sublattice-site coupling}
\label{sec:IIID}
The third configuration, labeled as AB, involves the $\Lambda$ system being coupled to both sublattice-sites A and B within the $x_1$
  unit cell, with coupling constants $g_1$ and $g_2$ respectively, and probing the upper energy band of the topological waveguide. The interaction Hamiltonian for this configuration is expressed as
\begin{equation}
    H_I^{AB} = g_1 a_{x_1}^{\dagger}\ket{g}\bra{e} + g_2 b_{x_1}^{\dagger}\ket{g}\bra{e} + \text{h.c.}
\end{equation}
Here, $a_{x_1}^\dagger$ and $b_{x_1}^\dagger$ represent a creation operator of a photon at A and B sublattice-site of $x_1$ unit cell. The total Hamiltonian of the system is: \(H_T^{AB} = H_{TW} + H_A + H_I^{AB}\). Using the secular equation: \(H_T^{AB}\ket{E_k} = \omega_k \ket{E_k}\), the equations of motion for the system are
\begin{subequations}
\renewcommand{\theequation}{\theparentequation\alph{equation}}
\begin{align}
\label{eq:23a}
 (\omega_e - \Delta_c) u_a + ({\Omega}/{2}) u_e = \omega_k u_a, \\
\label{eq:23b}
 g_1 u_A(x_1) + g_2 u_B(x_1) + \omega_e u_e + ({\Omega}/{2}) u_a = \omega_k u_e, \\
\label{eq:23c}
 -t_1 u_B(x_1) - t_2 u_B(x_1-1) + g_1 u_e = \omega_k u_A(x_1), \\
\label{eq:23d}
 -t_1 u_A(x_1) - t_2 u_A(x_1+1) + g_2 u_e = \omega_k u_B(x_1).
\end{align}
\end{subequations}
Solving the above Eqs. (\ref{eq:23a})-(\ref{eq:23d}) yields the scattering equations for A and B sublattice-site coupling, respectively
\begin{subequations}
\renewcommand{\theequation}{\theparentequation\alph{equation}}
\begin{align}
\begin{split}
\label{eq:24a}
-t_1 u_B(x_1) - t_2 u_B(x_1-1) = & (\omega_k - V_1) u_A(x_1) \\ 
& - V_2 u_B(x_1). 
\end{split}
\end{align}
\begin{align}
\begin{split}
\label{eq:24b}
 -t_1 u_A(x_1) - t_2 u_A(x_1+1) = & (\omega_k - V_3) u_B(x_1)\\ 
 & - V_2 u_A(x_1).
\end{split}
\end{align}
\end{subequations}
\noindent Here, the potentials \(V_1\), \(V_2\), and \(V_3\) are defined as \(V_1 = 4 g_1^2 G\), \(V_2 = 4 g_1 g_2 G\), and \(V_3 = 4 g_2^2 G\), where
$ G = ({\Delta_k + \Delta_c})/[{4 \Delta_k (\Delta_k + \Delta_c) - \Omega^2}]. $
Solving Eqs. (\ref{eq:24a}) and (\ref{eq:24b}) by using the single photon scattering formalism as discussed in Appendix leads to the transfer matrices corresponding to A and B sublattice-site coupling as
\begin{subequations}
\renewcommand{\theequation}{\theparentequation\alph{equation}}
\begin{align}
U_A = \begin{pmatrix}
        1 - \frac{(V_1 + V_2 e^{-i \phi})}{2i (t_1 - V_2) \sin{(\phi)}} & \frac{ -(V_1 + V_2 e^{i \phi}) e^{-i \theta}}{2i t_2 \sin{(\phi)}} \\
        \frac{(V_1 + V_2 e^{-i \phi}) e^{i \theta}}{2i t_2 \sin{(\phi)}} & 1 + \frac{(V_1 + V_2 e^{i \phi})}{2i (t_1 - V_2) \sin{(\phi)}}
    \end{pmatrix},
\end{align}
\begin{align}
U_B = \begin{pmatrix}
        1 + \frac{(V_3 + V_2 e^{i \phi})}{2i t_2 \sin{(k+\phi)}} & \frac{ (V_3 + V_2 e^{-i \phi}) e^{-2i kx_1}}{2i t_2 \sin{(k + \phi)}} \\
        -\frac{(V_3 + V_2 e^{i \phi}) e^{2i kx_1}}{2i t_2 \sin{(k + \phi)}} & 1 - \frac{(V_3 + V_2 e^{-i \phi})}{2i t_2 \sin{(k+\phi)}}
    \end{pmatrix}.
\end{align}
\end{subequations}
Here, $U_A$ and $U_B$ represent the transfer matrix corresponding to the A and B sublattice-site respectively.
In the expression for \(U_A\), the term \(\theta = 2(kx_1 + \phi)\). The total transfer matrix for this configuration is calculated as \(U_{AB} = U_B \cdot U_A\). By setting \(g_1 = g \alpha\) and \(g_2 = g (1 - \alpha)\) where $\alpha$ is a parameter that controls the amount of two couplings. we obtain the general expression for the transmittance. This expression can be applied to all configurations by adjusting the value of \(\alpha\) such that $A (\alpha = 1)$ and $B (\alpha =0)$ and its values vary as $0 < \alpha < 1$ for AB coupling. By using the similar approach as discussed in the above sections, the transmittance of AB configuration is calculated as
\begin{equation}
t_{AB} = \frac{2 i t_2 \sin{(k)}[t_1 - V \alpha (1 - \alpha)]}{2 i t_1 t_2 \sin{(k)} + V \omega_k A},
\end{equation}
where, $ A = [2 \alpha (1 - \alpha)(e^{-i \phi} - 1) + 1]$. The transmittance depends on the sign of \(\delta\), photon energy \(\omega_k\), the potential \(V\), and the parameter \(\alpha\). Therefore, the transmittance is topology-dependent and can be controlled by varying these parameters.\

\begin{figure}
    \centering
        \includegraphics[width= 1\linewidth]{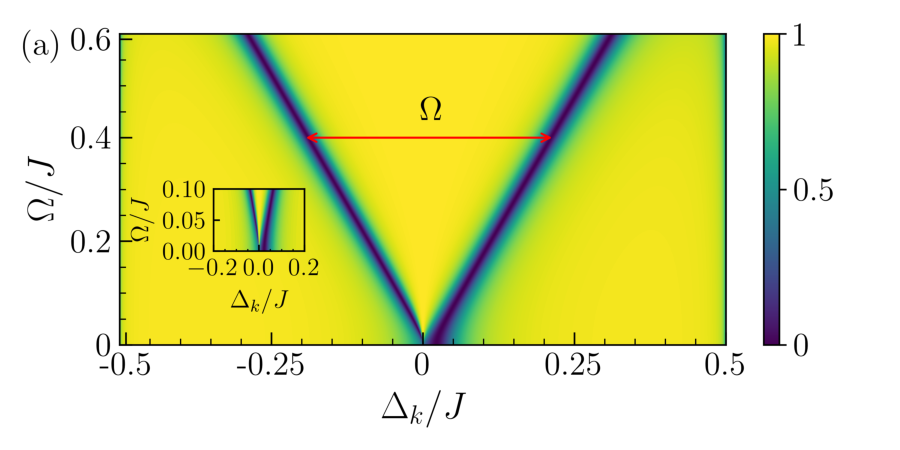}
        \includegraphics[width= 1\linewidth]{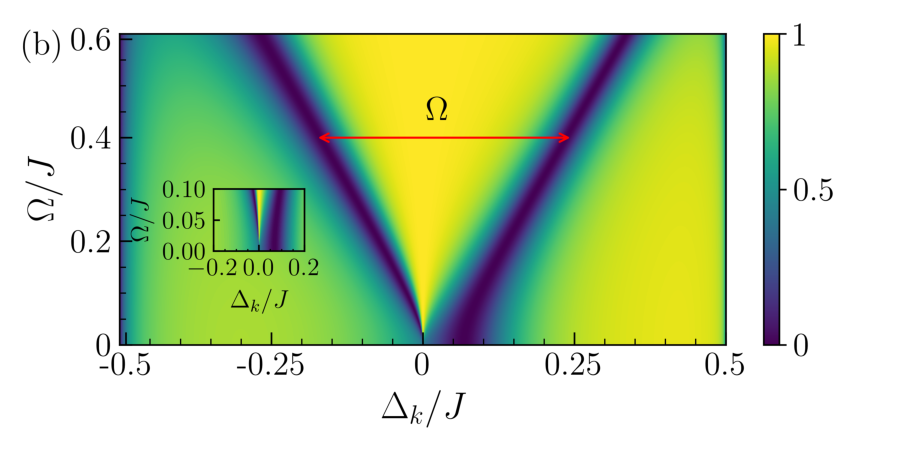}
    \caption{The 2D contour plot represents the transmission intensity \( T_{AB} = |t_{AB}|^2 \) as a function of the single photon detuning \(\Delta_k = (\omega_k - \omega_e)/J\) and the control field strength \(\Omega/J\). The coupling constant is fixed at \(g = 0.2J\) and the atomic transition frequency set to \(\omega_{eg} = \omega_e = 1.5J\). Subfigure (a) corresponds to the case where dimerization constant \(\delta > 0\) (\textit{i.e.}, \(\delta = 0.5\)), while subfigure (b) represents the case where \(\delta < 0\) (\textit{i.e.}, \(\delta = -0.5\)). The inset focuses on the transmission variation for small values of the Rabi frequency \(\Omega\).}
    \label{fig:contour_plot}
\end{figure}

Fig. \ref{fig:contour_plot} depicts the variation in transmission intensity $T_{AB} = |t_{AB}|^2$ as a function of the detuning of the incident photon $\Delta_k/J$ and the control field Rabi frequency $\Omega/J$ for $\delta > 0$ and $\delta<0$, respectively. The inset provides a closer view of the regimes in a small limit of $\Omega$. From the inset of Figs. \ref{fig:contour_plot}(a) and \ref{fig:contour_plot}(b), it can be observed that in the weak control field limit $\Omega \approx 0$, the point of zero transmission intensity is slightly shifted from $\Delta_k =0$. The reason for this shift will be discussed later. In an intermediate limit of $\Omega$, the regime showing minimum transmission intensity followed by maximum transmission intensity and then again followed by minimum intensity corresponds to the Fano line shape followed by the Lorentzian lineshape. This Fano line shape is much sharper for $\delta<0$ as compared with $\delta>0$ as shown in the insets.  As the Rabi frequency $\Omega$ increases, this Fano line shape transitions into the ATS regime.\

In the analysis of the system's response across different regimes of the control field Rabi frequency \(\Omega\), we examine the pole equation of the transmittance \(t_{AB}\) at \(\Delta_c = 0\). The two poles corresponding to the single excitation of the waveguide mode are expressed as
\begin{equation}
\Delta_k = \frac{i g^2 \omega_k A}{4 t_1 t_2 \sin{(k)}} \pm \sqrt{\frac{\Omega^2}{4} - \frac{g^4 \omega_k^2 A^2}{ 16 t_1^2 t_2^2 \sin^2{(k)}}}.
\end{equation}
\renewcommand{\figurename}{FIG.}
In the first regime, where \(|\Omega| \ll |g^2 \omega_k A / 2 t_1 t_2 \sin{(k)}|\), the given pole equation can be simplify to give two solutions \textit{i.e.}, \(\Delta_k = {i g^2 \omega_k A}/{2 t_1 t_2 \sin{(k)}}\) and \(\Delta_k = 0\). Solving for the first pole, we obtain
\begin{equation}
\begin{split}
    \Delta_k = & \frac{i g^2 \left[ 2 \alpha (1-\alpha) (-t_1 - t_2 \cos{(k)} - \omega_k) + \omega_k \right]}{2 t_1 t_2 \sin{(k)}} \\
    & + \frac{g^2 \alpha (1-\alpha)}{t_1},
\end{split}
\end{equation}

\noindent where the real part of \(\Delta_k\) corresponds to the position of the transmission dip. Notably, the dip does not occur exactly at \(\Delta_k = 0\) instead, it is shifted by \(\delta \omega_e = {g^2 \alpha (1-\alpha)}/{t_1}\), which is recognized as the Lamb shift \cite{bello2019unconventional}. This shift is induced by vacuum fluctuations in the waveguide and represents a radiative correction to the excited state of the atom. The magnitude of this shift depends on the coupling strength \(g\) and the parameter \(t_1 = J(1 + \delta)\), which is sensitive to the sign of \(\delta\). Consequently, the Lamb shift exhibits distinct behaviors for \(\delta> 0\) and \(\delta < 0\). 
It is important to note that the Lamb shift is present only in the AB configuration, whereas it is absent in the A and B configurations. This difference arises from the symmetric coupling of the emitter to the topological waveguide in the A and B configurations, resulting in either a zero or negligible density of states at the emitter frequency. As a consequence, virtual photon exchange is prevented, and no Lamb shift occurs. This can be further confirmed by setting \(\alpha = 0\) or \(\alpha = 1\), which yields \(\delta \omega_e = 0\). In contrast, the coupling in the AB configuration leads to a nonzero density of states, enabling the emitter to interact with the topological waveguide modes in a manner that induces a shift in its energy levels.\
\renewcommand{\figurename}{FIG.}
 \begin{figure*}
  \centering
\includegraphics[width=0.49\linewidth]{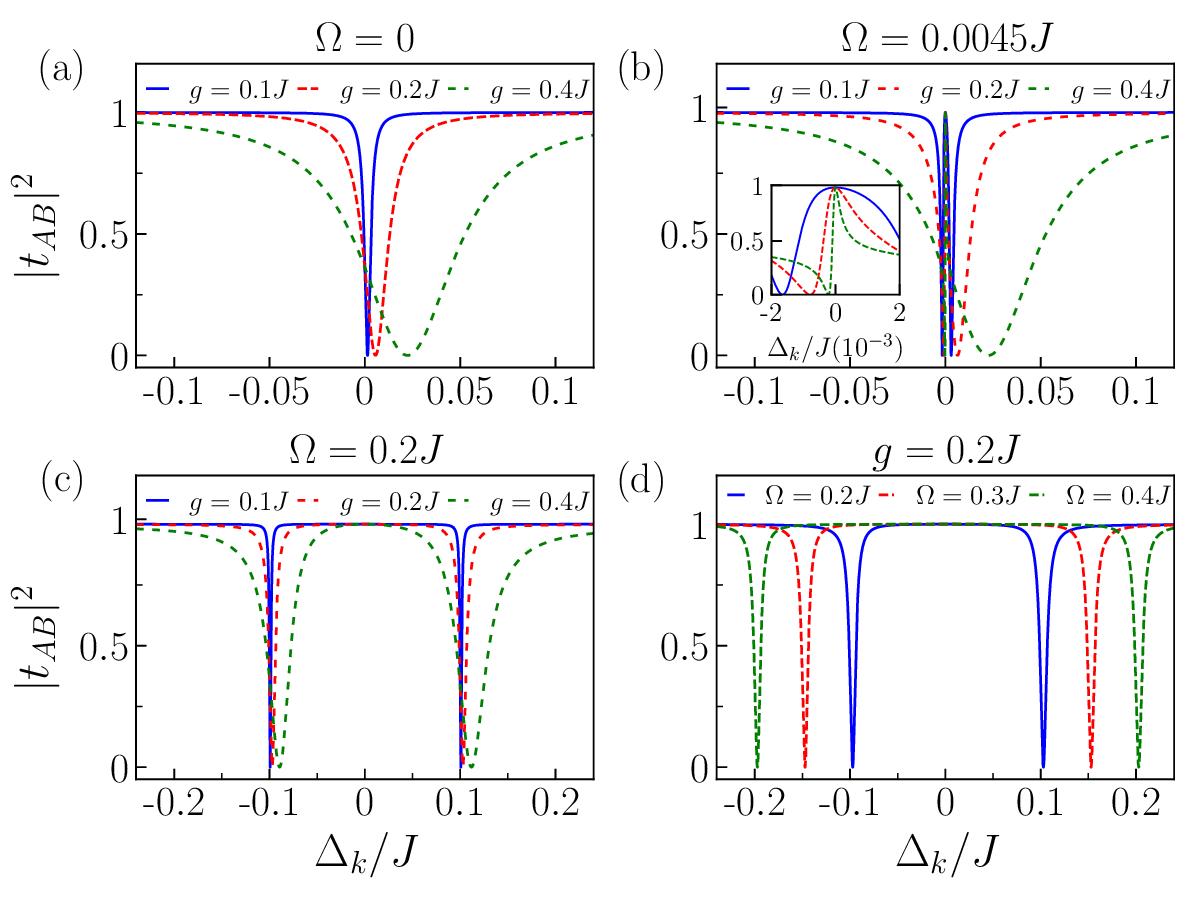}
\includegraphics[width=0.49\linewidth]{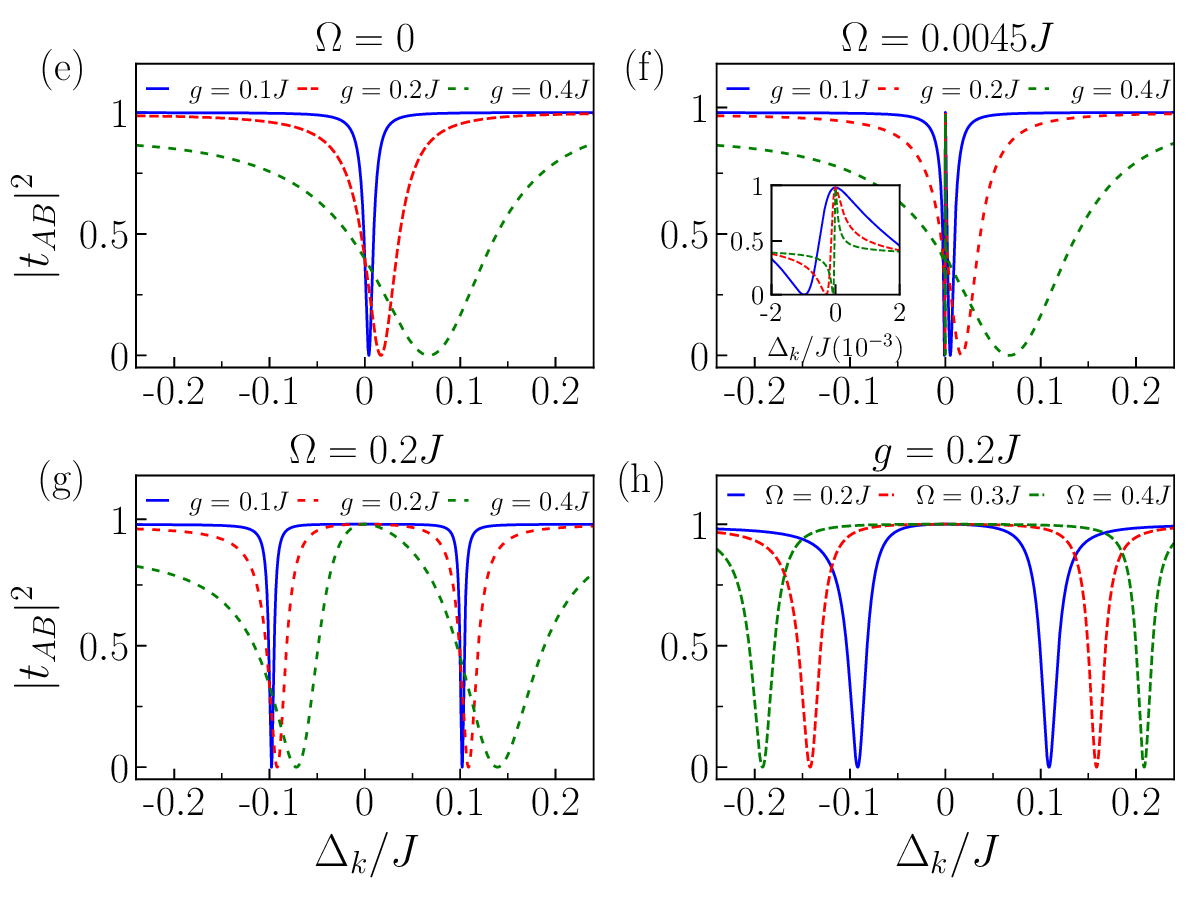}
  \caption{Transmission spectrum of a single photon as a function of the incident photon detuning, $\Delta_k = (\omega_k - \omega_e)/J$, in AB sublattice-site configuration.
  Subfigures (a)-(d) correspond to $\delta = 0.5$, while (e)-(h) correspond to $\delta = -0.5$, illustrating various values of the coupling constant $g$ and control field strength $\Omega$, with the atomic transition frequency fixed at $\omega_e = \omega_{eg} = 1.5J$.
(a) and (e) show the transmission spectrum in the absence of a control field $\Omega = 0$. (b) and (f) depict the transmission spectrum under a weak control field $\Omega = 0.0045J$. (c) and (g) present the transmission spectrum in the presence of a strong control field $\Omega = 0.2J$. In these subfigures, the solid blue, dotted red, and dotted green lines represent transmission for coupling constants $g = 0.1J$, $g = 0.2J$, and $g = 0.4J$, respectively. (d) and (h) shows the transmission spectrum for various control field strengths: $\Omega = 0.2J$ (blue), $\Omega = 0.3J$ (red), and $\Omega = 0.4J$ (green), with a fixed coupling constant of $g = 0.2J$. The inset zooms in the Fano line shape regime for a more detailed view.}
\label{fig:6}
\end{figure*}

In the second regime, where \( |\Omega| \approx |g^2 \omega_k A / 2 t_1 t_2 \sin(k)| \), the give pole equation simplifies to: $
\Delta_k = {i g^2 \omega_k A}/{4 t_1 t_2 \sin(k)}.$ This equation is indicative of EIT effect. 
In the third regime, where \( |\Omega| \gg |g^2 \omega_k A / 2 t_1 t_2 \sin(k)| \), the pole equation becomes: $\Delta_k = [{i g^2 \omega_k A}/{4 t_1 t_2 \sin(k)}] \pm {\Omega}/{2}.$
Expanding this expression, we obtain
\begin{align}
\nonumber
    \Delta_k  = & \frac{i g^2 \left[ 2 \alpha (1-\alpha) (-t_1 - t_2 \cos(k) - \omega_k) + \omega_k \right]}{4 t_1 t_2 \sin(k)} \\
    & + \frac{g^2 \alpha (1-\alpha)}{2 t_1} \pm \frac{\Omega}{2}.
\end{align}
This solution corresponds to the ATS. 
 Two dips appear in the transmission spectrum at the positions $[{g^2 \alpha (1-\alpha)}/{2 t_1}] \pm {\Omega}/{2}$. These features reflect the interaction of the system with a control field, revealing the characteristic signature of ATS.\

In this configuration, $t_{AB}$ is dependent on the sign of $\delta$, and the topology of the topological waveguide plays a significant role. Transmission intensity $T_{AB}= |t_{AB}|^2$ has been plotted for different signs of $\delta$, \textit{i.e.}, $\delta>0$ and $\delta<0$ in Figs. \ref{fig:6}(a)-\ref{fig:6}(d) and Figs. \ref{fig:6}(e)-\ref{fig:6}(h) respectively. Here, the system behavior is slightly different in both cases. From Figs. \ref{fig:6}(a) and \ref{fig:6}(e), we can see that in the absence of $\Omega$, the system behaves like a two-level system and has a Lorenzian shape with $R=1$. The amount of linewidth is controlled by $g$. The solid blue, dotted red, and dotted green line in Figs. \ref{fig:6}(a) and \ref{fig:6}(e) have been plotted for g = $0.1J$, $0.2J$, and $0.4J$ respectively. The transmission dip is not exactly at $\Delta_k =0$ it's shifted by an amount $\delta \omega_e$ \cite{bello2019unconventional}.\ 

In the presence of a control field, for $|\Omega| \approx | g^2\omega_k/2 t_1 t_2 \sin{(k)}|$ the spectrum gives an EIT response with a fano line shape. The reflection is followed by a sudden transmission. The reason of getting this fano line shape is the interaction of the discrete state of the $\Lambda$ system with a continuum formed by the combined modes of both
sublattices A and B of the topological waveguide. It appears in AB configuration only because of interference of scattering amplitude from different pathways of coupling with A and B sublattice-site \cite{Limonov2017}.
The Fano line shape corresponding to $\delta<0$ is named as the Topological Fano line shape and it is found to be much sharper than $\delta>0$ as shown in Figs. \ref{fig:6}(b) and \ref{fig:6}(f) \cite{PhysRevLett.122.014301}. This result is very similar to 2 two-level emitters coupled with a topological waveguide in AB configuration \cite{bello2019unconventional}. On further increasing the $\Omega$, the transparency window increases and the distance between two reflection dips also increases. The two dips are at the asymmetric position with respect to $\Delta_k = 0$ because of the lamb shift term. The one dip at $\Delta_k = g^2\alpha(1-\alpha)/2t_1 - \Omega/2$ and another at $\Delta_k = g^2\alpha(1-\alpha)/2t_1 +\Omega/2$ as shown in Figs. \ref{fig:6}(c) and \ref{fig:6}(g). The solid blue, dotted red, and 
 dotted green color lines have been plotted for different values of $g$. Keeping the $g$ fixed Figs. \ref{fig:6}(d) and \ref{fig:6}(h) have been plotted for different values of $\Omega$, which showing that splitting is getting control by $\Omega$.\

This configuration is topology dependent, $\delta<0$ representing the topologically robust regime, which is immune to small disorders and fabrication imperfections \cite{PhysRevLett.122.014301}. Operating in this configuration and achieving a sharper Topological Fano line shape would be advantageous for the design of various quantum devices. It holds potential for single photon switching applications, as the reflection and transmission probabilities exhibit rapid transitions from $T=0$ to $T=1$ by slightly changing the photon detuning. This makes it a promising candidate for efficient optical switches. Additionally, Fano resonances are highly sensitive to external perturbations, which opens up possibilities for utilizing this system in sensing applications.
\section{Experimental Implementation}
\label{sec:IV}
The proposed model system can be experimentally realized with the superconducting quantum circuits (SQCs) \cite{PhysRevX.11.011015}. The circuit analog of the 1D SSH waveguide can be implemented with the fabrication techniques for superconducting metamaterials. A $\Lambda$ system realized with a Josephson-junction-based Floxonium qubit \cite{PhysRevLett.120.150504, PhysRevLett.95.087001} can be coupled to a resonator site of the SSH waveguide.
 It can also be realized in photonic crystal cavities coupled with quantum dot and spin defects. The cavities can be designed with different diameters, which are equivalents to A and B resonators, and the distance between these two different size holes can be controlled to get the different conditions of $t_1$ and $t_2$ \cite{Gu2021}. 
 
\section{Conclusion and Discussion}
\label{sec:V}
We have investigated a single photon scattering under various configurations in a 1D topological waveguide coupled to a $\Lambda$ system. 
One of the dipole-allowed transitions in the $\Lambda$ system is coupled to the topological waveguide mode, while the other transition is driven by an external control field. 
By using the scattering formalism we derive an expression of transmittance and studied the single photon transmission spectrum at different control field strengths, for the A, B, and AB sublattice-site couplings.
 In the weak field regime of A and B sublattice-site configuration, a dip in the transmission spectrum is observed, centered at the atomic resonance frequency \( \omega_{eg} \). At this frequency, the input photon is completely reflected. However, as control field strength increases, an EIT emerges due to quantum interference of different pathways. Further increasing the control field strength, widens the transparency window and reaches the ATS regime. Thus, the transmission at the atomic resonance frequency can be switched from zero to one by tuning the control field strength. This tunability makes this configuration suitable for single photon switching devices.
In the AB sublattice-site configuration, the system can switch between trivial and non-trivial phases, depending on the sign of the dimerization constant. It leads to a topology-dependent transmission spectrum. In the weak control field regime, we observe perfect reflection with a Lorentzian line shape at the Lamb-shifted frequency. As the control field strength increases, a Fano line shape appears and shifts to ATS higher control field strength. Notably, the Fano line shape is significantly sharper for non-trivial case, which we refer to as a "Topological Fano line shape". Due to the sharpness of the Fano line shape, even a slight change in detuning can shift the transmission from zero to one, which can enable high-speed switching with good efficiency.
 In this non-trivial phase, the system is robust against small disorders within the system. Different configurations of this model can yield various desired outcomes and can be experimentally realized in systems such as superconducting circuits and photonic crystal cavities.

\section*{Appendix: Scattering Formalism for transfer matrix}
\label{sec:appendixA}
\appendix
\renewcommand{\figurename}{FIG.}
\begin{figure}
    \centering
\includegraphics[width=0.65\linewidth]{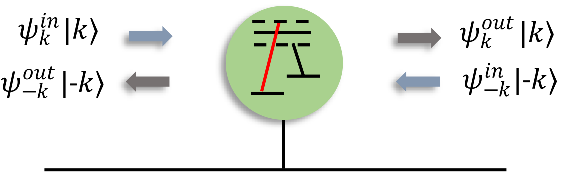}
    \caption{Schematic representation of a scattering eigenstate, with various amplitudes involved. The $\Lambda$ system separates the space into left and right regimes, with incoming modes (light blue) traveling toward the emitter and outgoing modes (brown) moving away from the emitter.}
    \label{fig:enter-label}
\end{figure}
Let us consider $\Lambda$ system is coupled to the sublattice-site A at $x_1$ unit cell of the topological waveguide. To solve the scattering equation, we used the following ansatz
\renewcommand{\theequation}{A\arabic{equation}} 
\setcounter{equation}{0} 
\begin{equation}
\ket{\Psi_k} = 
    \begin{cases} 
    \psi_k^{\text{in}}\ket{k}+ \psi_{-k}^{\text{out}}\ket{-k}, & \text{for } j < x_1, \\
    \psi_k^{\text{out}}\ket{k}+ \psi_{-k}^{\text{in}}\ket{-k}, & \text{for } j \geq x_1. 
    \end{cases} 
\end{equation}
Here, $\ket{\pm k} = u_{\pm k}^{\dagger}\ket{\text{vac}}$ or $\ket{\pm k} = l_{\pm k}^{\dagger}\ket{\text{vac}}$, depending on the energy band under consideration. 
The coefficients of the scattering eigenstate for the A sublattice-site coupling in position representation can be expressed as
\begin{subequations}
\renewcommand{\theequation}{\theparentequation\alph{equation}}
\begin{align}
    u_A(j) = \pm \begin{cases} \psi_k^{\text{in}}e^{i(kj+\phi_k)} + \psi_{-k}^{\text{out}}e^{-i(kj+\phi_k)}  & \text{$j \leq x_1$}\\ \psi_k^{\text{out}}e^{i(kj+\phi_k)} + \psi_{-k}^{\text{in}}e^{-i(kj+\phi_k)}  & \text{$j \geq x_1$}\\
    \end{cases},
\end{align}
\begin{align}
    u_B(j) = \begin{cases}
        \psi_k^{\text{in}}e^{i kj} + \psi_{-k}^{\text{out}}e^{-i kj} & \text{$j < x_1$}\\
        \psi_k^{\text{out}}e^{i kj} + \psi_{-k}^{\text{in}}e^{-i kj} & \text{$j \geq x_1$}\\
    \end{cases}.
\end{align}
\end{subequations}
Here, the $\pm$ sign in $u_A(j)$ expression refers to the upper and lower energy bands of the topological waveguide, respectively. The matching boundary condition at the coupling's position can be written as: $ u_A^-(x_1) = u_A^+(x_1).$
By putting $u_A(x_1), u_B(x_1)$ and applying the boundary condition in the scattering equation we can relate the amplitude on the left and right side of the emitter's position as
\begin{equation}
    \begin{pmatrix}
        \psi_k^{\text{out}}\\
        \psi_{-k}^{\text{in}}
    \end{pmatrix}
    = U \begin{pmatrix}
        \psi_k^{\text{in}}\\
        \psi_{-k}^{\text{out}}
    \end{pmatrix},
\end{equation}
where, $U$ represents the transfer matrix and is given by
\begin{equation}
    U = \begin{pmatrix}
        t_{11} & t_{12} \\
        t_{21} & t_{22}
    \end{pmatrix}.
\end{equation}
From the transfer matrix, we can compute the scattering matrix $S$, which relates
the incoming modes with the outgoing modes as
\begin{equation}
    \begin{pmatrix}
        \psi_k^{\text{out}}\\
        \psi_{-k}^{\text{out}}
    \end{pmatrix}
    = S \begin{pmatrix}
        \psi_k^{\text{in}}\\
        \psi_{-k}^{\text{in}}
    \end{pmatrix},
\end{equation}
with
\begin{equation}
    S = \begin{pmatrix}
        t_{11} - ({t_{12}t_{21}})/{t_{22}} & {t_{12}}/{t_{22}}\\
        
        -{t_{21}}/{t_{22}} & {1}/{t_{22}}
    \end{pmatrix} = 
    \begin{pmatrix}
        t_L & r_R\\
        r_L & t_R
    \end{pmatrix}.
\end{equation}
Here, $t_{ij}$ represents the transfer-matrix elements, while $t_{L/R}$ and $r_{L/R}$ represents the scattering- matrix elements. They correspond to the transmission and reflection probability amplitudes for a wave coming from the left/right. If the evolution is unitary, meaning there are no photon losses, the following relations hold:
\begin{equation}
    |t_L|^2 + |r_L|^2 = |t_R|^2 + |r_R|^2 = 1,
\end{equation}
and
\begin{equation}
    |t_L|^2 + |r_R|^2 = |t_R|^2 + |r_L|^2 = 1.
\end{equation}
Thus, we have the conditions \( |t_L| = |t_R| \) and \( |r_L| = |r_R| \). Furthermore, if the system has time-reversal symmetry ($H$ is real), as is the case in our model, the scattering is reciprocal, meaning that
\begin{equation}
    t_L = t_R.
\end{equation}

\bibliography{bibb} 

\end{document}